\documentstyle[aps,prl,epsf]{revtex}

\begin{document}

\twocolumn[\hsize\textwidth\columnwidth\hsize\csname
@twocolumnfalse\endcsname

\title{Thermal Contraction and Disordering of the Al(110) Surface}

\author{Nicola Marzari,$^{1,\dagger}$ David Vanderbilt,$^1$
Alessandro De Vita,$^{2,3}$ and M.\ C.\ Payne$^4$}
\address{$^1$Department of Physics and Astronomy, Rutgers University, 
Piscataway, New Jersey 08854-8019 \\
$^2$INFM and Department of Material Engineering and Applied Chemistry, 
University of Trieste, I-34149 Trieste, Italy\\
$^3$Institut Romand de Recherche Num\'{e}rique en Physique des
Mat\'{e}riaux, PPH-Ecublens, CH-1015 Lausanne, Switzerland\\
$^4$Cavendish Laboratory (TCM), University of Cambridge, Madingley
Road, Cambridge CB3 0HE, England}
\date{\today}

\maketitle

\begin{abstract}
Al(110) has been studied for temperatures up to 900~K via ensemble 
density-functional molecular dynamics. The strong anharmonicity displayed by
this surface results in a negative coefficient of thermal expansion, where the 
first interlayer distance decreases with increasing temperature. 
Very shallow channels of oscillation for the second-layer atoms in the 
direction perpendicular to the surface support this anomalous contraction, 
and provide a novel mechanism for the formation of adatom-vacancy pairs, 
preliminary to the disordering and premelting transition. Such 
characteristic behavior originates in the free-electron-gas bonding at a 
loosely packed surface.

\vspace*{2mm}
\end{abstract}

\vskip2pc]

\narrowtext

Metal surfaces exhibit a remarkable behavior as a function
of the temperature. Thermodynamic stability is often determined by a
delicate balance between energetic and entropic effects, and can
lead to a rich phenomenology for the phase diagrams of different systems.
Unreconstructed face-centered cubic (110) surfaces (e.g.\ Al, Cu, Ni) 
display a damped oscillatory pattern of interlayer relaxations,
starting with a large contraction between the first and
the second layer\cite{LEED,Ho85}.
Such behavior originates in the response of surface atoms
to under-coordination: moving towards the underlying layer,
they increase their surrounding charge density while reducing the 
corrugation of the surface and the lateral tensile strain\cite{strain}.
When the temperature is raised, this under-coordinated layer can 
start to disorder even before the melting temperature of the bulk 
is reached.
While the suggestion that a surface 
could act as a nucleation stage for melting had long been made,
experimental evidence of a reversible melting transition limited to
the outer surface layers came only recently\cite{premelting}.
For the case of Al(110), several experimental techniques 
(ion blocking and shadowing, electron or neutron diffraction, 
He scattering) have since shown
a clear onset of disordering at temperatures between 770 K and 
815 K\cite{premelting_al110}, whereas the bulk melting 
temperature is 933 K.
Computer simulations based on different models (effective-medium 
theory\cite{EMT,EMT2}, embedded-atom method\cite{EAM}, glue 
models\cite{glue}) have then been applied to the study of several
(110) surfaces (Pb, Al, Cu, Ni), and surface premelting was observed in
all cases.

However, many issues remain unresolved. Extensive low-energy electron
diffraction (LEED) studies in 
Al(110)\cite{LEED}(c) show a {\it negative} thermal expansion coefficient
for the first interlayer distance, and a large 
positive one (twice the bulk value) for the second interlayer distance.
These findings are
at variance with widely held general theoretical 
considerations\cite{jayanthi}, and with the results of
available computer simulations for Cu, Ni, and Al\cite{EMT2,EAM,glue}
which predict an expansion of the
first interlayer distance with temperature. In addition, model
calculations fail to reproduce the zero-temperature multilayer relaxation 
pattern\cite{EMT2,EAM}, predicting only
the contraction of the first interlayer.
On Al(110) the premelting transition is preceded by an anomalous 
proliferation of 
adatoms on the surface\cite{premelting_al110}, for which there is no
reliable microscopic picture. Finally, the degree of anharmonicity 
and anisotropy of the different surface layers, as opposed to the bulk, is 
not known, due to the experimental difficulty in resolving different layers.

Ab-initio molecular dynamics (MD) 
simulations of metal surfaces are
very challenging, and only few and limited studies have been attempted.
We use here an approach that we recently introduced\cite{edft,myphd}
(ensemble density-functional theory (eDFT)),
together with a technical improvement for the Brillouin Zone (BZ) 
integrations (so called ``cold smearing''), which 
is particularly suited to MD simulations. 
Applying this scheme to the case of Al(110), we provide the first
theoretical confirmation of a negative thermal expansion for this
surface, and excellent agreement with the experiments
for the temperature-dependent multilayer relaxations.
Moreover, we present a novel, and in retrospect simple, picture of the 
microscopic mechanisms that lead to this anomalous thermal contraction
and to the surface disordering associated with premelting.

In first-principles calculations for 
metals it is customary to introduce a fictitious electronic temperature 
$\sigma$\cite{sandrophd,broadening}, to broaden the density of states 
and to smooth the discontinuities at the Fermi energy $\mu$, greatly 
improving the sampling accuracy of a given set of k-points. 
It is very convenient to choose a broadening that has
zero first- and second-moments, so that the resulting
electronic free energy does not have any quadratic dependence on the 
broadening temperature, and neither do its derivatives with respect to 
any external parameter\cite{myphd,Methfessel89} 
(e.g. the Hellmann-Feynman forces, or the stress tensor).
In the existing schemes this is achieved at the price of allowing for 
{\it negative} orbital occupancies\cite{Methfessel89}, 
so that problems can arise in self-consistent calculations where the total 
electronic density may become negative. Here, we present 
a broadening scheme leading to an occupation function that is {\it 
positive} definite. 
Occupation broadening convolutes the density of states with a 
broadening of the $\delta$ function\cite{sandrophd,broadening};
the cold-smearing broadening is
\begin{equation}
 \tilde{\delta}(x)\,=\,\frac{2}{\sqrt{\pi}}\,
 \exp^{-\left(x-\frac{1}{\sqrt{2}}\right)^2}\,
 \left(2-\sqrt{2}\,x\right)\,\,.
\end{equation}
Spin-degeneracy is assumed here, and $x\,=\,\frac{\mu-\epsilon}{\sigma}$.
The ``generalized entropic functional'' $S\,=\,\sum_is_i$
that can be derived\cite{sandrophd}
and the occupation numbers
$f_i\,=\,\int_{-\infty}^{x_i}\,\tilde{\delta}(x)\,dx$
can all be expressed in terms of pseudoenergies 
$\tilde{\epsilon_i}$ \cite{Pederson91} ($x_i\,=\,
\frac{\mu-\tilde{\epsilon_i}}{\sigma}$); in particular
\begin{equation}
s_i\,=\,\frac{1}{\sqrt{\pi}}\,
 \exp^{-\left(x_i-\frac{1}{\sqrt{2}}\right)^2}\,
 \left(1-\sqrt{2}\,x_i\right)\,\,.
\end{equation}
No practical difficulty to the self-consistent calculations is caused by 
the fact that some spin-degenerate occupancies can still exceed 2; this was
also the case for the choices set forth in Ref. \onlinecite{Methfessel89}.

The calculations use the local-density approximation (LDA)
and norm-conserving pseudopotentials, with
a plane-wave basis cutoff of 11 Ry. The bulk properties of 
Al are well represented:
the lattice parameter is 3.96 (4.02) \AA, the elastic constants 
C$_{11}$=117 (114) GPa, 
C$_{12}$=66 (62) GPa, and C$_{44}$=39 (32) GPa (experimental results
at 0~K \cite{Kamm64} are in parenthesis).
The simulation cell is a $3\times 3$ 8-layer
Al(110) slab, containing 72 atoms separated by 8.5 \AA\
of vacuum.  k-point sampling is performed
with the $\frac{1}{4},
\frac{1}{4},\frac{1}{4}$ Baldereschi point, using $\sigma$=0.5 eV of 
cold smearing. 

The zero-temperature structural properties
are summarized in Table \ref{tab:cell}:
good and consistent agreement with the experimental results is registered.
The 8-layer calculation has been performed using the same finite
cell and sampling of the MD simulations; this 
introduces some small finite-size errors, that can be evaluated exactly at 0~K 
comparing them with a fully converged calculation (a $1\times 1$ 15-layer slab
with $12\times 12\times 2$ k-point sampling). 

Constant-temperature MD simulations have been performed 
using a Gaussian thermostat and a
leapfrog velocity Verlet algorithm to integrate the 
ionic equations of motion\cite{thermostat}, using a timestep of 8 fs.
A set tolerance for each timestep of 5
meV/cell in the spread of the total energies over the 
last 5 electronic iterations 
resulted in a negligible drift of the constant of motion 
(less than 1.5 meV/atom/ps for a microcanonical run).
The lattice parameter parallel to the surface was fixed
applying the experimental thermal expansion coefficient for the 
bulk to the LDA equilibrium lattice parameter. 
We followed five runs,
at increasing temperatures of 400, 600, 700, 800
and 900~K, for 5, 10, 6, 6, and 6 ps respectively.

\begin{figure} 
\epsfxsize=3.0 truein
\epsfbox{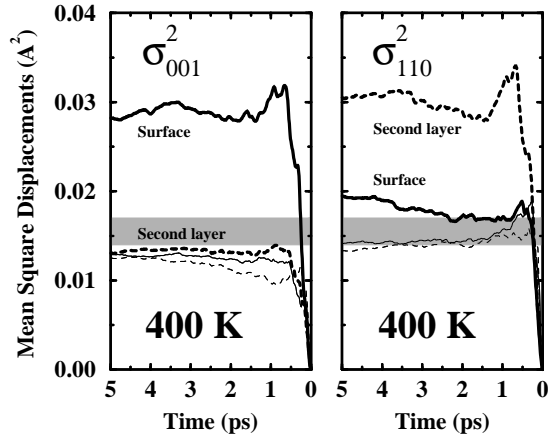}
\caption{Layer-resolved mean square displacements at 400~K; left panel
shows the [001] components (henceforth labeled $x$), the right panel 
the [110] ones ($z$). The third- and fourth-layer data are given by the 
unlabeled solid and dashed thin lines. The shaded area corresponds to the bulk 
experimental values of Ref.~\protect\cite{msd}. The horizontal axis shows the time
over which averaging was performed.}
\label{fig:rms.400} 
\end{figure} 
We present in Fig.~\ref{fig:rms.400} our results 
for the mean square displacements (MSDs)
in the different layers, from the surface 
to the interior of the crystal, during a 5 ps run
at 400~K.
The horizontal scale shows the decremental time: the plot starts
with the averages over the full run,
then proceeds by discarding a progressively longer initial segment. This 
approach highlights the initial thermalization time (negligible in this
case),
the flatness of the plateau for the converged time average, and provides
an estimate for the statistical errors. The left panel of
Fig.~\ref{fig:rms.400} shows the $[001]$ component for the MSDs
(we label it $x$); this component is parallel 
to the surface and perpendicular to the $[\overline{1}10]$ rows that 
characterize the $(110)$ surface. The right panel shows the $[110]$
$z$ component, perpendicular to the surface. 
The time averages are
well converged, with the third- and fourth-layer results very close to each
other (giving us confidence on the absence of finite-size effects), 
and close to the experimental bulk values\cite{msd}. 

Two results stand out from the simulation. First, the MSDs
in the $x$ direction are twice as large for the surface atoms than
for those in all the other inner layers. While it can be expected that the
undercoordinated atoms on the surface should be more loosely bound,
the large difference with the averages for the lower layers is notable.
Second, the MSDs in the $z$ direction (i.e.\ perpendicular to the surface) 
are  {\it much larger in the second layer than in the first layer}.
This is a distinctive feature of this crystallographic orientation
that was first encountered in embedded-atom simulations of Ni(110) and 
Cu(110)\cite{EAM}. In Al the effect is more striking due to its
free-electron-gas behavior. A simple rationalization can be offered:
since the (110) surface is very open, atoms in the second 
layer have natural channels of oscillation perpendicular 
\begin{figure}
\epsfxsize=3.0 truein
\epsfbox{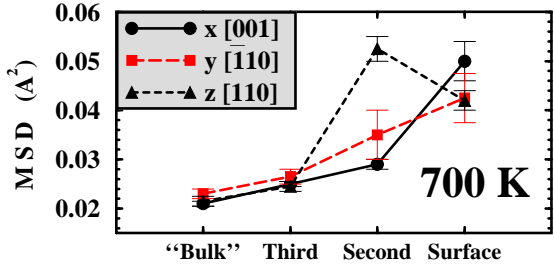}
\caption{Layer-resolved mean square displacements at 700~K. The $x,y$ ones
parallel to the surface are in solid black and dashed grey 
($x$ is perpendicular and $y$ is 
parallel to the $[\overline{1}10]$ rows); the $z$ ones, 
perpendicular to the surface 
are in dashed black. ``Bulk'' refers to the two inner layers.}
\label{fig:rms.700} 
\end{figure} 
to the surface and directed towards the vacuum. 
The charge density on the top of the second-layer atoms is
still quite homogeneous, and the bonds are easily stretched, leaving 
thus the freedom for the atoms to move back and forth along these
channels. On the other hand, atoms in the first (surface) layer see the
vacuum acting as a hard wall, limiting their mobility outwards;
their largest oscillations are thus parallel to the surface and 
perpendicular to the $[\overline{1}10]$ rows. 

\begin{figure} 
\epsfxsize=3.0 truein
\epsfbox{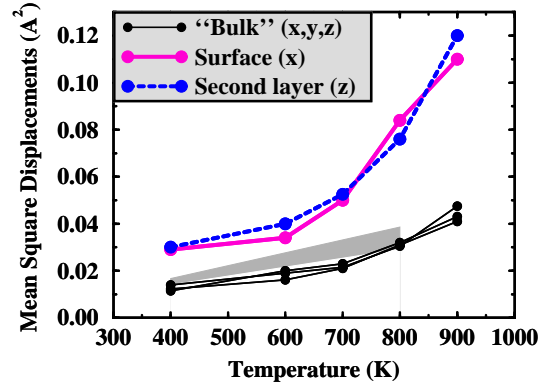}
\caption{Layer-resolved mean square displacements as a function of 
temperature.  The shaded area corresponds to the bulk experimental values of 
Ref.~\protect\cite{msd}.}
\label{fig:rms_t} 
\end{figure}

\begin{figure} 
\epsfxsize=2.8 truein
\epsfbox{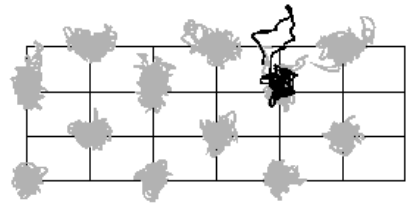}
\caption{Projected view of the MD trajectories in the
$(\overline{1}10)$ plane, for the top 4 layers of the slab. 
The large excursion towards the surface of a second-layer atom
has been highlighted.}
\label{fig:adatom} 
\end{figure}

The anisotropic behavior of the surface dynamics can be gauged by looking 
at Fig.~\ref{fig:rms.700}, where the MSDs at 700~K are plotted in all three
crystallographic orientations as a function of the
layer depth. Moving from the bulk to the surface, one can observe
that the third layer still behaves in a bulk-like fashion: 
the MSDs are isotropic, and they are only slightly larger
than those in the two layers below. 
The anisotropy becomes very distinct in the second layer, with its
characteristic large MSDs perpendicular to the surface, and persists in
the first layer, for which the `easy' channels are parallel to the surface
and across the close-packed $[\overline{1}10]$ rows. Some of the
components for the MSDs in the first and second layers can be up to 2-3 
times their bulk counterparts. These enhancements
near the surface are due to the lower coordination;
in addition, a higher degree of 
anharmonicity makes these surface MSDs increase much more rapidly with 
temperature than the bulk ones. 
This becomes apparent from the plot in
Fig.~\ref{fig:rms_t} of the MSDs as a function of the temperature.
The innermost layers show 
isotropic MSDs, with some deviation from the linear regime only above 700~K
(in the harmonic regime the MSDs dependence on the temperature is exactly 
linear). The outer layers, on the contrary, are strongly anharmonic.
The very large increases in the vibrational amplitudes along the 
`easy' channels are precursors to the creation of adatoms and vacancies 
on the surface, that lead to the disordering and premelting of the surface.
In fact, we observe that with increasing temperature atoms in the
second-layer start making increasingly large slow excursions towards the
surface. One of these events is shown in Fig.~\ref{fig:adatom}; the 
highlighted atom temporarily pops out from the surface. In another
event the second-layer atom remained outside the surface, creating 
an adatom-vacancy pair where the vacancy is initially
in the second layer (this void is quickly
filled up by a surface atom). In this second case
the adatom diffused away via exchange diffusion.

\begin{figure} 
\epsfxsize=3.0 truein
\epsfbox{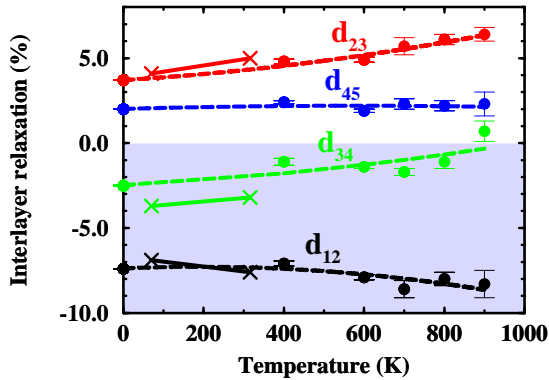}
\caption{Interlayer relaxations as a function of temperature.
Solid circles: eDFT MD simulations, with their statistical error bars; 
dashed lines: quadratic least-squares fit; crosses: experimental LEED values
(Ref.~\protect\cite{LEED}(c)). \
The eDFT data have had their 0~K finite-size corrections added rigidly 
at all temperatures.}
\label{fig:interlayer} 
\end{figure}

The microscopic dynamics provides a clear explanation 
of the behavior of this surface, that displays an increasing
contraction of the first interlayer distance with temperature,
where a large expansion would have been expected. 
The contraction can be understood looking at the motion
of the second-layer atoms along these channels that are shallower
towards the vacuum. With increasing temperature, the center of mass
of the second layer moves outwards, since it is
not hampered by nearest-neighbors directly on top (the first layer is
staggered with respect to the second).  The first layer is more limited 
in its expansion, since the vacuum acts as a hard wall. The end result is that 
the average distance between the first and the second layer decreases with
temperature. This decrease is then offset by a larger thermal increase 
of the distance between the second and the third layer.
The results of our simulations for the interlayer relaxation as a function of 
temperature (see Fig.~\ref{fig:interlayer}) are in very good 
quantitative agreement with the LEED data (Ref.~\onlinecite{LEED}(c)). 

In conclusion, our calculations on Al(110) represent the first
extensive first-principles molecular-dynamics simulations of
the dynamics on a metal surface, presenting both an insightful picture
of the microscopic dynamics and a remarkable agreement with the
available experimental data.
The microscopic dynamics of this surface is peculiar, and governed
by the interplay between the free-electron-gas behavior of the bulk and the 
quasi-covalent bonding of the undercoordinated surface atoms. Two
distinct soft channels of oscillation have been identified.
One channel is at the surface in the [001] direction, perpendicular to 
the close-packed surface grooves. The other, unexpected, is
perpendicular to the surface but confined to the second-layer atoms.
It is this channel that is responsible for the observed anomalous contraction 
of the surface with temperature. Additionally, it provides a novel, 
favored mechanism for the generation of adatom-vacancy pairs, whose 
proliferation is precursor to the disordering and premelting transition.

N.M.\ acknowledges support by NSF grant DMR-96-13648;
these calculations have been performed on the Hitachi S3600 
at the University of Cambridge High Performance Computing Facility.

$^\dagger$ Present address: Center for Computational 
Materials Science, Naval Research Laboratory, Washington DC.

\begin{table}
\caption{Theoretical and experimental values for the interlayer relaxations
in Al(110); theoretical values are at 0~K.}
\begin{tabular}{lcccc} 
& d$_{12}$ & d$_{23}$ & d$_{34}$ & d$_{45}$ \\  \tableline
LDA, 8 layers & -6.1 \% & +5.5 \% & -2.2 \% & +1.7 \% \\
LDA, 15 layers & -7.4 \% & +3.8 \% & -2.5 \% & +2.0 \% \\ 
LEED$^a$ (100~K) &-8.6 \% & +5.0 \% & -1.6 \% & \\
LEED$^b$ (70~K) 
&-6.9 \% & +4.1 \% & -3.7 \% & +1.7 \% \\ 
\end{tabular}
$^a$Reference \onlinecite{LEED}(a).\\
$^b$Reference \onlinecite{LEED}(c).
\label{tab:cell}
\end{table}

\end{document}